\documentclass[12pt,prd,aps,amssymb,amsmath,tightenlines]{revtex4}
\usepackage{graphics}
\usepackage[pdftex]{graphicx}
\begin{document}


\title{Meta-CTA Trading Strategies based on the Kelly Criterion}

\author{ Bernhard K. Meister}

\affiliation{ Department of Physics, Renmin University of China, Beijing, China 100872}
\date{\today }

\begin{abstract}


\noindent The impact of Kelly-optimizing portfolio managers on the price process 
 is explored with the help of a simple model. Investments are restricted  to one risky and one  riskless asset.
The Kelly optimizers invest under the assumption that the risky asset follows a geometric Brownian motion with fixed risk premium and volatility. They keep a fixed proportion  called the leverage ratio, defined as the ratio of the risk premium and the volatility, in the risky asset. This requires   regular adjustment of the portfolio weights as prices evolve. The price impact relating trading activity and price change is assumed to be given by a power law. These two rules are combined and lead to a deterministic price dynamic. Price changes provide feedback to the Kelly optimizers, who adjust their portfolios. This in turn leads to further price changes and closes the loop. For different choices of the model parameters, most importantly the leverage ratio and the power law coefficient of the price impact, one gets qualitatively  different dynamics. The results can be expressed as a phase diagram. For different combinations of the two parameters one gets three  types of qualitatively different behaviour.   In the first phase, the change of the price gradually decreases to zero. In the second phase the sign of the price change switches at each step. The third phase is most interesting, since the change of the price increases at each step and leads to a run-away effect.  This analysis is useful for understanding  the impact of Commodity Trading Advisors (CTA) on the underlying price process, and is of increasing relevance, since the assets under management in this strategy have ballooned.  As a result,  slippage becomes an issue and the profitabilities of conventional CTA strategies are affected. In contrast,  meta-CTA strategies can be devised to exploit the predictability inherent in the toy model dynamics by avoiding critical areas of the phase diagram or by taking a contrarian position at an opportune time.

\end{abstract}
\maketitle


\section{Introduction}
\noindent The aim of this paper is to develop a simple model to explore some particular consequences of Kelly optimizers on the price process. This is for example useful to understand some of the side-effects of the explosive  growth of so-called Commodity Trading Advisers (CTA) over the last few decades.
A range of different investment styles are subsumed under the name Commodity Trading Advisers (Burghardt \& Walls 2011, Lemprire {\it et al.} 2014).
Such strategies   have existed for many years and invest in a variety of assets. For the purpose of this paper we focus on funds that base trading decisions on computer models. Human intervention, except when severe short-term market dislocations occur, is frowned upon. In particular we are interested in funds that try to exploit trend-based strategies, implemented through liquid financial instruments such as futures. 
The whole CTA space according to BarclaysHedge grew from \$300 million in 1980 to more than \$300 billion in 2014. Out of these assets around \$275 billion are estimated to be controlled by ``systematic traders". This group matches more or less the subset we are considering, since systematic funds, in contrast to discretionary ones, base trading decisions solely on computer models. 
These quantitative models often  rely on momentum or mean-reversion signals, i.e.~are concerned with trend following or trend reversion. Leverage is employed and funds can take long as well as short positions. All the usual liquid financial assets are employed, with a preferences for assets such as futures that allow naturally for long and short positions.
Holding periods vary, but are mostly longer than just intra-day, unlike high-frequency based statistical arbitrage strategies.  
The  toy model  that we shall construct can be used to show how CTA trading can impact the underlying instruments and effect their performance. In addition it will be shown in a heuristic way how this insight can be used to modify and improve trading strategies.

Market micro-structure, except where it touches on general price impact, will be ignored in what follows. Portfolio selection paired with risk control plays a prominent role for CTA managers, and the hope is that the results we describe are of practical relevance to the investment policies of portfolio managers. In particular, our approach may allow for the construction of meta-strategies that are able to  avoid the slippage  associated with the conventional crowded  CTA strategies, and in the optimal case will lead to predatory  trading strategies that exploit the predictability of conventional CTA strategies. The individual systematic CTA strategies are black boxes not known to outsiders. Nevertheless there seems to be a significant overlap between different CTAs. In addition, their long-running and well-documented success has led to some leakage of their ideas into the public domain.

The remainder of the paper is organised as follows. After a general description of the toy model in the next subsection, the two main components called Kelly criterion and price impact are briefly explored. 
In the subsequent section implications of the model are developed and a phase diagram produced. It will be shown how different values of the main parameters of price impact and the proportion of portfolio value in the risky asset will lead to qualitatively  different behaviour.  Next, the key assumptions are introduced  to gain a qualitative understanding of how Kelly optimizers effect  the underlying market.

\section{Market Impact and Kelly Criterion}

\noindent The  toy model at the centre of our approach and laid out below is based on two elementary assumptions. The model is constructed in such a way as to capture a set of stylized facts. The assumptions are related to portfolio construction and adjustment as well as trading impact. The result is a deterministic price dynamic, as the updating of positions, e.g.~end-of-day, leads  through price impact to a modification of price, which in the next cycle leads to an adjustment of the positions. This feedback loop continues indefinitely.

The two assumptions  are as follows. 
(i) The Kelly criterion (i.e.,~the optimal growth strategy) governs the portfolio selection.  Past stock behaviour is used as a guide to estimate future performance parameters of drift, volatility and correlation for a standard CTA-strategy, which ignores price impact. The use of information theoretic ideas in portfolio optimization goes back to Kelly (1956) and was further developed both in theory and practice by Thorp (1969). For a detailed review see MacLean {\it et al.} (2011). (ii)
The relation between investment amount and price impact is given by  a power law 
relating price change to investment amount, in other words, a relation of the form 
\begin{eqnarray}
\left(\Delta S \right)^{\gamma} \sim \Delta N,
\end{eqnarray}
where the asset price is given by $S$, and the number of shares in the asset  is given by $N$. This is a generally accepted relationship, i.e.~a stylized fact, and can be found the work of econophysicists such as Stanley, Farmer, and others. The Kelly criterion and price impact are combined to up-date in regular time intervals, e.g.~daily, the portfolio allocation in a self-contained way and produce a price dynamic for the underlying assets.

These key concepts  are sufficient to develop a tractable deterministic model. First the concepts are turned into equations, which can be analyzed and lead to a phase diagram relating parameter choices to price behaviour. The stability properties of the resulting price process can be deduced.

The market impact $\gamma$ 
is normally a number around $0.5$ 
 depending on asset class and country with a vast phenomenological literature to support the estimates.  It signifies how price impact scales with amount of directed buy or sell activity. The necessary liquidity is provided by dedicated market makers and nowadays to some extent also by High Frequency Traders. 
In the next paragraphs the Kelly criterion is further explored.

For the portfolio construction we rely on the Kelly criterion
introduced by Kelly (1956) in the wake of earlier information-theoretic results by Shannon to find the optimal betting amount in games with fixed known odds. It was later extended to the field of financial investments by Thorp (1969) and others. The  strategy  is equivalent to myopic log-utility maximization. In the process the entropy of the value process is maximized. Entropy up to a multiplicative constant can be uniquely characterized  by a small set of rules. For a modern derivation from three axioms based on information loss see Baez {\it et al.} (2012). 
The question, why investors should choose to maximize the log wealth, has a simple answer: according to Breiman's theorem (Breiman, 1960), the strategy gives the asymptotically optimal pay-out and dominates any other strategy, i.e in the long run trounces any different strategy. 
The Kelly criterion tells us for example that the optimal betting fraction is given by $p-q$, if a gambler is faced with a bet, where the probability to double the money is $p$ and to lose the initial stake is $q$ $(p>q, p+q=1)$. 
 
   The original idea has been extended to the general continuous time framework with $N$ arbitrarily correlated assets. The sensitivity to estimation errors in both expected drift and correlation has been studied and sensitivities, like the greeks of option theory, can be calculated. A general analysis of the Kelly criterion for the Ornstein-Uhlenbeck process can be found in Lv \& Meister (2009, 2010).
   
The Kelly criterion is not a panacea and has it critics:  see, in particular, Samuelson (1969, 1979). It is often regarded as too risky for practical investment, since (a) the chance of losing $ \epsilon$ is $1-\epsilon$ for $ \epsilon \in (0,1)$, (b) parameters are unstable, and (c) the sensitivities are hard to estimate in practice. One way of going beyond standard entropy is to consider the larger class of Renyi entropies, which provide a range of portfolios associated with a continuous risk aversion parameter. 

In the continuous case with one risky  asset with price process $\{S_t\}_{t\geq0}$ assumed by the Kelly-optimizers to evolve according to a geometric Brownian motion (GBM) with drift $\mu_t$ and volatility $\sigma_t$, 
\begin{eqnarray}
\frac{dS_t}{S_t}= \mu_t dt + \sigma_t dW_t,
\end{eqnarray}
and the riskless asset increasing with the short rate $r_t$ given by
\begin{eqnarray}
\frac{dB_t}{B_t}= r_t  dt,
\end{eqnarray}
with $B_0=1$, the optimal portfolio, with value process  $\{V_t\}_{t\geq0}$, maximizes the exponential growth. The result is given in terms of $\Lambda_t$, which is the proportion of the available assets invested in the risky asset at time $t$. We further set the drift $\mu_t$ equal to $r_t + \sigma_t \lambda_t $, where $\lambda_t$ is the market price of risk.
The value process of the portfolio has to satisfy
\begin{eqnarray}
\theta_t S_t + \phi_t B_t = V_t,
\end{eqnarray}
where $\theta_t$ is the number of shares in the portfolio, and $\phi_t$ is the number of units in the money market account.
In addition, the self-financing equation is
\begin{eqnarray}
\theta_t dS_t + \phi_t dB_t = dV_t.
\end{eqnarray}
These  relations allow us to calculate the optimal amount $\theta_t$. First we evaluate the differential of the growth of the value process
\begin{eqnarray}
d \log(V_t) &=& \frac{dV_t}{V_t} -\frac{1}{2} \frac{(dV_t)^2}{V_t^2}\nonumber\\
&=&  \frac{\theta_t dS_t + \phi_t dB_t } {V_t} - \frac{1}{2} \frac{(dS_t)^2}{V_t^2}\nonumber\\
&=& r_tdt+ \Big(\frac{\lambda_t \sigma_t \theta_t S_t}{V_t}  -\frac{1}{2}  \frac{\sigma_t^2 \theta_t^2 S^2_t}{V_t^2}\Big)dt +\frac{\theta_t S_t \sigma_t}{V_t} dW_t. 
\end{eqnarray}
The drift of the value process $\{V_t\}_{t\geq 0}$ is maximized by taking the derivative with respect to the freely chosen number of shares $\theta_t$, to obtain
\begin{eqnarray}
\theta_t^*:= \frac{\lambda_t}{\sigma_t} \frac{V_t}{S_t}, \nonumber
\end{eqnarray}
and  as a consequence
\begin{eqnarray}
\phi_t^*:= \Big(1-\frac{\lambda_t}{\sigma_t} \Big)\frac{V_t}{B_t}, \nonumber
\end{eqnarray}
It can be rewritten, and provides the definition of the leverage ratio 
\begin{eqnarray}
\Lambda_t^*: = \frac{\lambda_t}{\sigma_t}=\theta_t^* \frac{S_t}{V_t}.
\end{eqnarray}
This  derivation
holds for a general class of adapted processes for the drift, volatility, interest rate and market price of risk. The proportion of assets in shares is equal to the ratio of the market price of risk, which is equal to the Sharpe ratio, and the volatility, and can be rewritten as  $(\mu_t - r_t)/\sigma_t^2$. In subsequent sections we assume the leverage ratio, the risky asset price volatility, the interest rate and the market price of risk to be slowly changing variables. As a consequence, we assume the CTA investors to take these quantities  to be constant and not to be readjusted with the same speed as the portfolio weights.
In other words, for the time-scale we are interested in the price changes of the underlying asset as well as the connected portfolio weights are significant, but the parameters are held constant.

One could ask, if  there is an elegant way of implementing this optimal portfolio using derivatives. Indeed, such a derivative exist. Let us look at  the European call option in the Black-Scholes model with constant volatility and interest rate. 
The value of a call with strike $K$ is given by 
\begin{eqnarray}
C(S_t,K, \sigma, r,T-t)= S_t N(d_1)- K e^{-r(T-t)}N(d_2), 
\end{eqnarray}
where $N(\cdot)$ denotes the  normal cumulative distribution function, and $d_1= d_2+ \sigma \sqrt{T-t}= \frac{1}{\sigma  \sqrt{T-t} }\Big(\log(S_t/K_t) + (r+\sigma^2/2)(T-t)\Big)$.
The value process of the call can be replicated by a combination of stock and bond. The stock weight is given by, what is commonly called $\Delta$, $N(d_1)$ and the bond weight is given by $-K N(d_2)\exp(-r T)$. If  these weights coincide with the optimal leverage in stocks and bonds as calculated before, then   the portfolio with value $C(S_t,K,\sigma,r,T-t)$  achieves the stated purpose. This 
can be further elaborated by noting that for 
\begin{eqnarray}
N(d_1)&=&\frac{\lambda}{\sigma} \frac{C(S_t,K, \sigma, r, T-t)}{S_t}\nonumber\\
N(d_2)&=&-\Big(1-\frac{\lambda}{\sigma}\Big) \frac{C(S_t,K, \sigma, r,T-t)}{K} e^{r(T-t)}
\end{eqnarray}
 equivalence is achieved. 
  There is not always a single call available to reproduce the optimal  portfolio.  If for example $\lambda\leq \sigma$, then there is no solution. On the other hand, due to market completeness derivatives can never outperform the optimal stock plus bond portfolio. 
 In addition, since $N(d_1)\geq N(d_2)$, we have the further constraint
\begin{eqnarray}
\frac{\lambda}{\sigma} \frac{K}{S_t} \Big( e^{-r(T-t)}-1 \Big)\geq -1. \nonumber
\end{eqnarray} 
One can determine numerically combinations of  $K$ and $T-t$ for fixed  $S_t$, $\sigma$, $r$ and $\lambda_t$ that satisfy the constraints.
As time progresses the position would shift in strike or maturity. A special cases is treated in the Appendix.


The risky asset does not have to follow a geometric Brownian motion, but can instead belong to the class of geometric Levy models (GLM)  of the form
\begin{eqnarray}
\hat{S}_t= \hat{S}_0 e^{ r t + R(\lambda,\sigma) t + \sigma X_t - \psi(\sigma)t }
\end{eqnarray}
with  the expectation value 
$\langle e^{\sigma X_t} \rangle= e^{\psi(\sigma)t}$;
for   background and details see Brody {\it et al.} (2012).  where $\sigma,\,\, \lambda \,\,\& \,\,r$ are chosen to be constants, $\psi$ is called the Levy exponent, and $R(\lambda, \sigma)$ has the role of a risk premium. 
In this case the self-financing condition containing differentials is not directly applicable and instead we use the expansion of the logarithm $\log(1+x)= x - x^2/2+O(x^3)$.
The value process in the GLM case is 
\begin{eqnarray}
 \hat{\theta} \hat{S_t} +  \hat{\phi} B_t =  \hat{V_t}.
\end{eqnarray}
and for computational convenience we set $ \hat{V_0}=1$ and $B_0=1$.
We maximize again the logarithmic   utility
\begin{eqnarray}
 \langle  \log( \hat{\theta} \hat{S_t} + (1- \hat{\theta} \hat{S_0} )B_t )\rangle 
\end{eqnarray}
over all $ \hat{\theta}$.
 The logarithmic utility function in the short time limit, relevant for the optimisation, since the portfolio can be adjusted at any time, becomes
\begin{eqnarray}
 \langle  \log(  \hat{\theta} \hat{S_t} + (1-  \hat{\theta} \hat{S_0} )B_t )\rangle - \log( \hat{V_0})
 &=& \Big\langle \log\Big( 1+ \frac{ \hat{\theta} \hat{S_t} + (1-  \hat{\theta} \hat{S_0} )B_t - \hat{V_0}}{ \hat{V_0}}\Big)\Big\rangle\nonumber\\
 &=&\Big\langle ( \hat{V_t}-1)  - ( \hat{V_t}-1)^2/2 + \sum_{k=2}^{\infty}(-1)^k \frac{(V_t-1)^k}{k}
 \Big\rangle .\nonumber
\end{eqnarray}
The first term in the expansion is
\begin{eqnarray}
  \hat{V_t}-1  =   \hat{\theta} \hat{S_0} \Big( (1+ r t +R(\lambda, \sigma) t) e^{(\sigma X_t -\psi(\sigma) t)} -(1+ r t )\Big)+ r t +O(t^2),\nonumber\\
  \end{eqnarray}
and the expectation value of this term is
\begin{eqnarray}
\langle  \hat{V_t}-1 \rangle &=&   \hat{\theta} \hat{S_0}\Big\langle \exp( \sigma X_t - t \psi(\sigma))\Big\rangle(1+r t +R(\lambda,\sigma)t) - \hat{\theta} \hat{S_0}(1+rt)+ rt+  O(t^2)\nonumber \\
&=&   \hat{\theta} \hat{S_0} R(\lambda, \sigma) t + r t + O(t^2).
\end{eqnarray}
The expectation value of the second term is
\begin{eqnarray}
\langle ( \hat{V_t}-1)^2 \rangle = 
\hat{\theta}^2 \hat{S}^2_0 [t \psi(2 \sigma) - 2 t \psi(\sigma)]+ O(t^2),
\end{eqnarray}
while higher order terms can be ignored, since they are of order $O(t^2)$.
The combination of  terms gives
\begin{eqnarray}
 \langle  \log(\hat{V_t})\rangle= \hat{\theta}  \hat{S_0}\Big( R(\lambda,\sigma)t \Big)+rt-\frac{1}{2}\Big(\hat{\theta}  \hat{S_0}\Big)^2\Big(t \psi(2\sigma)-2 t \psi(\sigma)\Big)+ O(t^2).
 \end{eqnarray}
The optimal leverage ratio for GLM in the limit of $t\rightarrow 0$ is 
\begin{eqnarray}
\hat{\Lambda}= \frac{R(\lambda, \sigma)  
}{ \psi(2 \sigma)-2 \psi(\sigma)}.
\end{eqnarray}
As an example, for geometric Brownian motion  with $R(\lambda, \sigma) = \lambda\, \sigma$, $\psi(\sigma)= \sigma^2/2$, and the process $\{X_t\}_{t\geq 0}$ set equal to $\{W_t\}_{t\geq 0}$ one reproduces  the result derived above. The optimal ratio can also be calculated for jump diffusion and other popular models. Sensitivities of the optimal ratio to the various quantities appearing in the formula  like $\sigma$ and $\lambda$, as done in Lv {\it et al.} (2009 \& 2010), can be determined. 

In the next section for a  specific  price process and under some  additional suitable simplifications the dynamics  of the optimal value  as well as the underlying asset process are described. This leads to a deterministic dynamic with an associated phase diagram.

\section{The resulting  portfolio dynamic in  the one risky asset case}


\noindent The portfolio dynamics can be studied in a variety of ways. Here we look at  a particular simple case, where both the market price of risk and volatility estimate, $\lambda$ and $\sigma$, for the asset $\{S_t\}_{t\geq0}$ are assumed to remain unchanged and as a consequence the leverage ratio is also a fixed quantity.  In addition, the interest rate $r$ is taken to be constant. These  three quantities are considered  in this section as externally fixed parameters, i.e estimated by the Kelly investors either solely  from a fixed set of historical data or also based on forward looking data only occasionally updated incorporating future earnings estimates or the company growth rate as the case could be for equities.  
One could claim that this is not entirely consistent, since the leverage ratio is only true in a probabilistic setting, but the model we consider here is deterministic. This apparent paradox can be resolved by assuming that in this toy model  the portfolio investors 
are not aware of the deterministic nature of the price process and believe the risky asset to be governed by a geometric Brownian motion with fixed drift and volatility. 

The portfolio value  of a Kelly investor is then given by
\begin{eqnarray}
V_t= \frac{\lambda}{\sigma}V_t +\Big(1-\frac{\lambda}{\sigma}\Big)V_t,  
\end{eqnarray}
where 
$\lambda/\sigma$ 
is the optimal leverage ratio $\Lambda$ derived above.
 The change of the portfolio value due to the self-financing condition is given by
\begin{eqnarray}
dV_t= \frac{\lambda}{\sigma} V_t \frac{dS_t}{S_t} +\Big(1-\frac{\lambda}{\sigma}\Big)V_t \frac{dB_t}{B_t}
\end{eqnarray}
and
\begin{eqnarray}
\frac{dV_t}{V_t}= \Lambda  \frac{dS_t}{S_t} +\Big(1-\Lambda\Big) \frac{dB_t}{B_t}.
\end{eqnarray}
The change in the number of shares 
in the portfolio is given by
\begin{eqnarray}
 \Delta \theta_t
& =&\Lambda \frac{V_t + dV_t}{S_t+dS_t} - \Lambda \frac{V_t}{S_t},  \nonumber\\
 &=&\Lambda \frac{V_t}{S_t} \Big(\frac{dV_t}{V_t}-\frac{dS_t}{S_t}- \sigma^2 dt +O(dS_t^3)\Big). 
\end{eqnarray}
After some simplifications the combination of the last two equations leads to 
\begin{eqnarray}
 \Delta \theta_t
 \propto \Big[\frac{dV_t}{V_t}- \frac{dS_t}{S_t} \Big]
\sim \Big[(\Lambda -1 ) \frac{dS_t}{S_t}+ (1-\Lambda) r dt \Big].
\end{eqnarray}
We can further assume $rdt\sim 0$, since to neglect the change of the value of the money market account seems reasonable for short time intervals, in particular in the current interest rate environment. The last equation can then be rewritten as
\begin{eqnarray}
\Delta  \theta_t
\sim \Big[(\Lambda -1 ) \frac{dS_t}{S_t} \Big].
\end{eqnarray}
  For example, we therefore have
\begin{eqnarray}  
   \Delta  
\theta_t    \sim\frac{dV_t}{V_t}  -  \frac{dS_t}{S_t} \geq 0,  {\rm if}\,\,\,  dS_t>0 \,\,\,\,\& \,\,\,\, \Lambda >1. 
\end{eqnarray}

As a result, there are two conditions in this dynamical system that govern the changes in the number of shares held\begin{eqnarray}
 \Delta  \theta_t
 &\sim&\Big[(\Lambda -1 ) \frac{dS_t}{S_t} \Big] ,  \,\,\,\,\,\,\,\,\,\,\,\,\,\,\,\,\,\,\,\,\,\, \,\,\,\,\,\,\,\,\,\,\, \,\,\,\,\,\,\,\,\,\,\,\,\,\,\,\,\,\,\,\,\,\, \,\,\,\,\, \,\,\,\,\,\,\,\,\,\,\, \,\,\,\,\,\,\,\,\,\,\, \,\,\,\,\,\,\,\,\,\,\, \,\,\,\,\, \,\,\,\,\,\,\,\,\,\,\,  \,\,\,\,\,\,\,\,\,\,\, *
 \\
  \Delta \theta_t
  &\sim& \Big(\frac{dS_t}{S_t} \Big)^{\gamma}, \,\,\,\,\,\,\,\,\,\,\,\,\,\,\,\,\,\,\,\,\,\, \,\,\,\,\,\,\,\,\,\,\,  \,\,\,\,\,\,\,\,\,\,\, \,\,\,\,\,\,\,\,\,\,\, \,\,\,\,\,\,\,\,\,\,\, \,\,\,\,\,\,\,\,\,\,\, \,\,\,\,\,\,\,\,\,\,\, \,\,\,\,\,\,\,\,\,\,\, \,\,\,\,\,\,\,\,\,\,\, \,\,\,\,\,\,\,\,\,\,\, \,\,\,\,\,\,\,\,\,\,\, ** 
\end{eqnarray}
which are applied alternately.
\begin{figure}[th]
\begin{center}
  \includegraphics[scale=0.50]{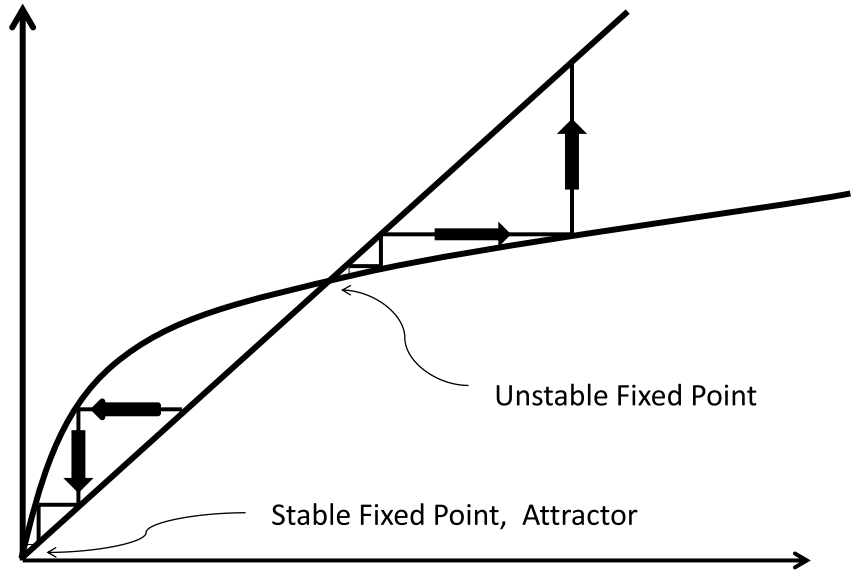}
  \caption{\textrm{Price-change versus position-size-change diagram. The $x$-axis is $dS_t/S_t$ and the $y$-axis corresponds to the change of the number of shares in the portfolio at each adjustment step.  The straight line with the slope dependent on $(\Lambda-1)$ relates price change to portfolio change and the square-root function relates change in number of shares to price impact. As a consequence the market drifts either to the stable fixed point at the origin or, if it starts to the right of the unstable fixed point, away to infinity.}.
  \label{fig:1}
  }
\end{center}
\end{figure}

These two relations are sufficient to already develop some intuition about the dynamics of the price process and the related portfolio changes as shown in Figure 1.
The size of $\Lambda$ governs the slope of the straight line from the origin. If $\Lambda$ is larger one then the number of shares added at each time stage, assuming the right sign for the slow changing variables like the drift and volatility, is positive.
The number is either ever increasing as is the case to the right of the unstable fixed point, where the two curves intersect, or is ever decreasing and
approaching zero as the process moves towards the left fixed point.
The analysis above holds true, if we assume all the `slow' parameters like $r, \lambda \& \sigma$ are fixed in time. A  realistic dynamic adjustment of these parameters, i.e. considering them as time dependent variables, plus unavoidable noise leads to more complex behaviour more closely aligned with what one can see in the financial market. This is not surprising, since the added layer of complexity allows a fine tuning of the system. 

The resulting dynamics can be most easily described graphically, and some additional insights can be garnered from Figure 1, where the dynamics of the price process vis-a-vis the portfolio changes are
given by arrows. One see that the portfolio as well as the asset price are  adjusted to bounce between the two curves. The upper curve relates the
position adjustment to price changes by the price impact equation governed by $\gamma$ which is normally around $0.5$. The lower curve relates the portfolio adjustment based on  the leverage ratio $\Lambda$ to the asset price change.
The leverage ratio is in each step modified to maximize the growth of the portfolio. As a result of the dynamics, the market drifts either to the stable fixed point at the origin or, if it starts to the right of the unstable fixed point, away to infinity.


\begin{figure}[th]
\begin{center}
  \includegraphics[scale=0.50]{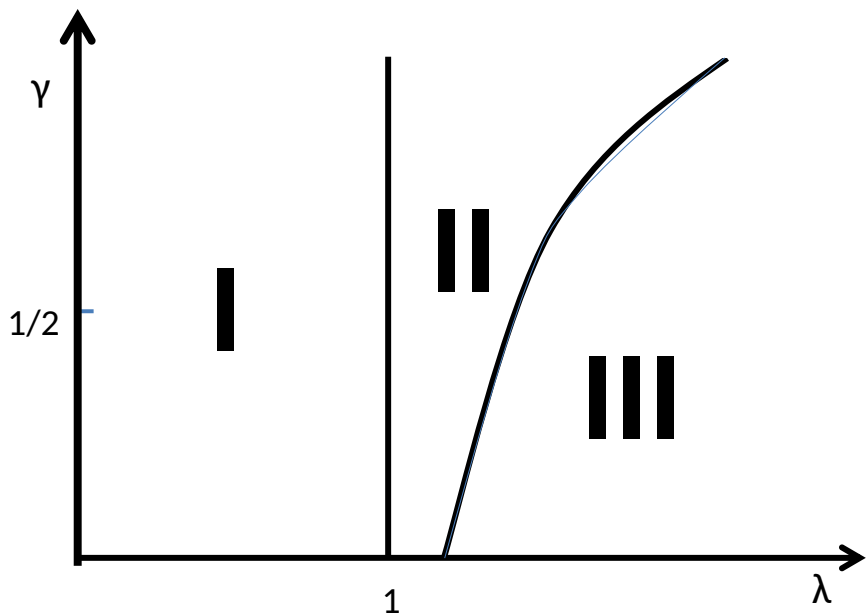}
  \caption{\textrm{Phase diagram. The phase diagram relates for a fixed initial change of  position size the price impact coefficient $\gamma$ around the value of $0.5$ to the leverage ratio $\Lambda$ on the x-axis. The diagram falls into three regions. In the region I the price process oscillates around zero and moves in the deterministic case towards zero. In the region II the price process monotonically decreases and tends towards zero. In region III the price process monotonically increases and the situation is unstable.  }.
  \label{fig:2}
  }
\end{center}
\end{figure}

In Figure 2 a rough phase diagram is displayed around the value of $\gamma$ of $0.5$ representing the $y$-axis and the leverage ratio $\Lambda$ the $x$-axis. It shows the existence
of three phases. Phase I represent the region, where the price change oscillates between positive and negative at each step and price changes get progressively smaller. Not displayed is the region, for larger values of $\gamma$, where the oscillations between positive and negative values get bigger with time and an instability arises.
The Phase II represents the region, where the changes of the price in each time step always have the same sign, but decrease with time.
In Phase III the price process explodes as price changes always have the same sign and get bigger with each time step. The portfolio adjustment increases also with time and the free float is eventually unable to accommodate the liquidity requirements. Explosive growth is unsustainable and leads to a market break-down and eventually to a dramatic price correction.
In the next section strategies are briefly described to exploit this predictability.

\section{Meta-CTA strategies}
\noindent A natural extension of the analysis of the earlier sections is to consider, how  other market participants can exploit rigid CTA strategies and how CTA managers can make their strategies more flexible. This leads to meta-CTA strategies.

The first step in the construction of a meta-CTA strategy is to determine, in which of the three phase the market is situated. If  the process is placed squarely in Phase III, then caution is required, since the simplified price process without noise will reach an instability due to larger and larger movements, and a major reversion is to be expected.
A meta-CTA investor should as a consequence deviate from the rigid CTA strategy  and reduce exposure, whereas a predatory market participant should take a position possibly through derivatives like vanilla options to exploit the eventual price reversion accompanied by a spike in volatility. A fast moving market participant might also try to exploit first the run-up of the asset and then the eventual correction using some well-defined rules, whose suitability might be profitably checked against historical data.

The other two Phases provide less direct opportunities. 
If the market is in Phase II, a move towards an equilibrium is expected. This suggests the absence of a large  reversion as well as a decline of volatility, which could be exploited through the selling of gamma.
If the market is in Phase I, volatility is likely to decline and the progressively decreasing price swings lead to a lack of directional opportunities.

As just described, out of the three phases, the most substantial opportunities exist in Phase III, where a run-away effect creates potentially an exploitable instability. These phenomenological ideas can be further 
buttressed by the study of market data and by including noise in the theoretical model to enable sensitivity analysis.

\section{Conclusion}
\noindent  The toy model studied in the paper shows what CTA managers are confronted with, if they rigidly adhere to a fixed view of the underlying price process.

The weakest assumption  of the paper, namely that the Kelly based investors misjudge the market
by assuming it to be described by a known geometric Brownian motion, is maybe less debilitating than it might initially appear, since the amount of data required to reasonably estimate the asset drift  is not available in markets with regular paradigm shifts. As a consequence, a  geometric Brownian motion model with fixed parameters for the risky asset is often taken as a starting point in finance.

What might seem surprising is the ability to create interesting price behaviour and portfolio dynamics without the addition of extraneous noise, i.e. in a deterministic setting. Of course, noise as well as a sensitivity analysis with regards to the various variables employed would make the model more realistic and useful. This will be studied separately.

It is comforting that even such a simple model already provides some food for thought and raises some practical questions. Is it for example
possible to determine in some heuristic way in which part of the phase diagram a stock  or other asset currently hovers?
Once the phase is determined one can adjust ones investment behaviour, e.g. Phase II has lower volatility and allows higher leverage, whereas Phase III requires a more cautious approach. 
This has to be compared with the volatility and drift assumptions for the original CTA model to show how a Meta-CTA strategy differs from the un-reflected version. It would be also of interest to see how the dynamics change, if one links the slow changing variables like asset drift and volatility to recent changes of the asset and adds noise.
Another questions of interest is how the model can be used to discover points where trends are reversed, i.e. how trend reversal is a natural aspect of the   unstable Phase III.

Another point of interest would be to link the approach described to the predictable rebalancing of leveraged exchanged traded funds (ETF), which in some Asian markets have a significant market share and are responsible for the majority of ETF trading activity. The link between leveraged ETF rebalancing  and equity volatility was already explored in Shum {\it et al.} (2015) and it would be interesting to see, how the methodology introduced in this paper could distinguish between different phases. 



  Helpful comments  by D.C. Brody and L.P. Hughston, including the  derivation in Section II of the leverage ratio $\Lambda_t^*$ based on the self-financing criterion, which is an  improvement to  the calculations one normally finds in the literature, are gratefully acknowledged.

\nocite{*}

\vspace{1cm}
\hspace{-0cm}{\bf APPENDIX:  AN EXAMPLE OF A CALL OPTION AS A KELLY OPTIMAL PORTFOLIO}

\noindent  Let us find the at-the-money call option $C$ with $r=0$  and positive market price of risk  $\lambda \geq 0$, 
such that the value of the Call option matches the optimal portfolio. Two constraints have to be matched. The stock  and bond weights of the replicating portfolio of the option have to coincide  with the  weights  calculated with the help of  the  Kelly criterion.
  By fixing the various quantities, the equations containing the normal cumulative distribution functions $N(d_1)$ and $N(d_2)$   can be combined and simplify  to 
\begin{eqnarray}
N(d_1)&=& \frac{1}{2-\frac{\sigma}{\lambda}}
\end{eqnarray}  
  using $N(d_1)=1-N(-d_1)=N(-d_2)$.
  There are always solutions for $N(d_1)$ to satisfy the equation above, if $ \lambda/ \sigma > 1 $.  The  function $N(d_1)$ lies in this case only in the interval $[1/2,1]$, since both $\lambda$ and $\sigma$ are positive. 
If  the combination $\sigma^*$ and $T^*-t^*$ is a solution, 
then so is $\sigma^* /  \alpha$ and $(T^*-t^*)\alpha^2$ for any positive $\alpha$. 


\end{document}